\input{aipcheck}

\documentclass[
    ,final            
  ]
  {aipproc}

\usepackage{amsmath}
\usepackage{amsfonts}
\usepackage{amsthm}

\def\tr{{\rm tr}}

\newcommand{\MeV}{\,{\rm MeV}}

\newcommand{\SU}{{\rm SU}}

\newcommand{\vx}{{\mathbf{x}}}
\newcommand{\vp}{{\mathbf{p}}}

\newcommand{\ignore}[1]{}

\newcommand{\cL}{{\cal L}}

\newcommand{\diag}{{\textrm{diag}}}

\layoutstyle{8x11single}

\begin{document}

\title{Polyakov loop, Hadron Resonance Gas Model and Thermodynamics of QCD~\thanks{Invited plenary talk given by E.~Meg\'{\i}as at the XXXVI Brazilian Workshop on Nuclear Physics, 1-5 September 2013, Maresias, S\~ao Sebasti\~ao, S\~ao Paulo, Brazil.}}


\author{E.~Meg\'{\i}as}{
  address={Grup de F\'{\i}sica Te\`orica and IFAE, Departament de F\'{\i}sica, Universitat Aut\`onoma de Barcelona, Bellaterra E-08193 Barcelona, Spain}
}

\author{E.~Ruiz Arriola}{
  address={Departamento de F{\'\i}sica At\'omica, Molecular y Nuclear and Instituto Carlos I de F{\'\i}sica Te\'orica y Computacional, Universidad de Granada, E-18071 Granada, Spain}
}

\author{L.L.~Salcedo}{
  address={Departamento de F{\'\i}sica At\'omica, Molecular y Nuclear and Instituto Carlos I de F{\'\i}sica Te\'orica y Computacional, Universidad de Granada, E-18071 Granada, Spain}
}

\begin{abstract}
We summarize recent results on the hadron resonance gas description of QCD. In
particular, we apply this approach to describe the equation of state and the
vacuum expectation value of the Polyakov loop in several representations.
Ambiguities related to exactly which states should be included are discussed.
\end{abstract}

\maketitle


\section{Introduction}
\label{intro}

The thermodynamics of Quantum Chromodynamics (QCD) has received much
attention due to the possible existence of new phases of nuclear
matter at sufficiently high temperatures and densities, such as the
quark-gluon plasma~\cite{MeyerOrtmanns:1996ea}. The phase transition
from a confined to a deconfined phase is characterized by a change of
order parameters connected with different symmetries of the
system. The relevant degrees of freedom in the deconfined phase are
quarks and gluons, while in the confined phase these constituents form
color singlet bound states and mostly resonances, i.e. hadrons and
possibly glueballs.  When physical quark masses are considered, recent
lattice simulations indicate that this transition is actually a
hadron--quark-gluon crossover~\cite{Aoki:2006we}.

The Hadron Resonance Gas (HRG) model is a useful approach to describe
the thermodynamics of QCD in the confined phase. It is based on the
assumption that physical observables in this phase admit a
representation in terms of hadronic states which are treated as
non-interacting and point-like particles~\cite{Hagedorn:1984hz}. These
states are usually taken as the conventional hadrons listed in the
review by the Particle Data Group (PDG)~\cite{Beringer:1900zz}, so the
completeness of the PDG states is assumed. A commonly used order
parameter for the hadron--quark-gluon crossover is the Polyakov loop
in the fundamental representation, as it is related to the free energy
of a heavy quark ``$h$'' placed in a thermal
medium~\cite{Borsanyi:2010bp,Lucini:2012gg,Bazavov:2013yv},
\begin{equation}
L_{\bf 3} = \langle \tr \, {\sf P} e^{i g \int_0^{1/T} \! A_0\, dx_0 } \rangle 
\simeq e^{-F_h/T}
\,,
\label{eq:PL}
\end{equation}
where ${\sf P}$ indicates path ordering and $A_0$ is the Euclidean time
component of the gluon field. While the HRG model has been traditionally
applied to study the equation of state of QCD, a similar hadronic
representation for the Polyakov loop in the fundamental representation was
recently formulated in~\cite{Megias:2012kb}. This approach has been confronted
with recent lattice data and confirmed the accuracy of the model for temperatures
in the range $150\,\MeV < T <
190\,\MeV$~\cite{Megias:2012kb,Bazavov:2013yv}. The Polyakov loop in any other
representation can be computed in lattice QCD, and the existence of Casimir
scaling relations has been guessed from lowest order perturbation theory~\cite{Dumitru:2003hp,Gupta:2007ax}. The
generalization of the hadronic representation to other representations besides
the fundamental one is a natural step forward which could be used to test the
existence of exotic states in the QCD spectrum. In this communication we 
briefly overview the HRG model, and study
its realization within a particular quark model. We 
advance some results for the Polyakov loop in the representations ${\bf 8}$, ${\bf 10}$, ${\bf 15}$ and ${\bf 27}$.

\section{QCD at Finite Temperature and Symmetries}
\label{sec:symmetries}

QCD is accepted nowadays as the fundamental theory of strong interactions.  The QCD Lagrangian
\begin{equation}
{\cal L}_{\rm QCD} = \frac{1}{2} \tr \left( G_{\mu\nu}^2\right) + \sum_f \overline{q}_f (i \gamma_{\mu} D_{\mu} - m_f) q_f \,, \qquad  G_{\mu\nu} = \partial_\mu A_\nu - \partial_\nu A_\mu - ig[A_\mu,A_\nu] \,, \label{eq:Lagrangian_qcd}
\end{equation}
describes the dynamics of quarks $q$ and gluons $A_\mu$, which are the fundamental fields in the theory. ${\cal L}_{\rm QCD}$ is constructed to be invariant under color gauge transformations.
The QCD equation of state can be derived from the partition function
\begin{equation}
Z_{\rm QCD} = {\rm Tr}\, e^{-H/T} = \int {\cal D} A_{\mu ,a} \exp \left[- \frac1{4} \int d^4 x (G_{\mu\nu}^a)^2 \right] {\rm Det} (i \gamma_{\mu} D_{\mu} - m_f) \,. \label{eq:Z_qcd}
\end{equation} 
Within the imaginary time formalism of finite temperature field theory, in the
path integral of Eq.~(\ref{eq:Z_qcd}) the bosonic and fermionic fields obey
periodic and anti-periodic boundary conditions in the Euclidean time,
respectively, with periodicity $\beta=1/T$. These boundary conditions are
preserved under periodic gauge transformations.

In the massless quarks limit $m_f = 0$, the theory is invariant at
the classical level under scale transformations.
This symmetry is broken by quantum corrections through the inevitable
regularization, leading to a non-zero value for the interaction measure or
trace anomaly $\epsilon-3P$. In this limit, ${\cal L}_{\rm QCD}$~is also
invariant under chiral transformations, which are spontaneously broken by the
chiral condensate in the vacuum $\langle {\bar q}q \rangle \ne 0$. At high
enough temperature, i.e. for $T > T_\chi$, the chiral symmetry is restored,
and this is signaled by a vanishing value of $\langle {\bar q}q \rangle$.

In the opposite limit of infinitely heavy quarks $m_f \to \infty$,
there is a larger symmetry related to the center ${\mathbb Z}(N_c)$ of
the gauge group, as gauge transformations which are periodic in
Euclidean time modulo an element of the center are allowed,
i.e. $\Lambda(\vec{x},x_0+\beta) = z \, \Lambda(\vec{x},x_0)$,
$z^{N_c} = 1$. In the Polyakov gauge $\partial_0 A_0 = 0$, an example
of such a transformation is $\Lambda(x_0) = e^{i2\pi
x_0 \lambda/(N_c \beta)}$, where $\lambda =
\diag(1,1,\ldots,1-N_c)$,
under which the Polyakov loop transforms as
\begin{equation}
L_{\bf 3} = \langle \tr_c \, e^{i g A_0/T} \rangle \longrightarrow  e^{i2\pi/N_c}  L_{\bf 3} \,. \label{eq:transf_LT}
\end{equation}
From Eq.~(\ref{eq:transf_LT}) it follows that $L_{\bf 3}$ vanishes in
the ${\mathbb Z}(N_c)$ symmetric phase. This means that the heavy
quark free energy diverges, corresponding then to a confined
phase. Likewise a non ${\mathbb Z}(N_c)$ symmetric phase is
characterized by a non-vanishing value of the Polyakov loop, leading
to finite free energy which corresponds to a deconfined phase.  So the
Polyakov loop can be used as an order parameter for the
confinement-deconfinement phase transition in QCD in the limit
$m_f\to\infty$.  The picture that emerges is that in QCD there are two
phase transitions: one related to chiral symmetry, and another one
related to confinement (or center symmetry), with transition
temperatures $T_\chi$ and $T_D$ respectively, which are quark-mass
dependent. In the real world, because of finite quark masses, the
chiral condensate and the Polyakov loop are approximate order
parameters only, and lattice simulations predict $T_\chi \approx
T_D \approx
200 \,\textrm{MeV}$~\cite{Borsanyi:2010bp,Bazavov:2009zn}. The
physical mechanism behind the close agreement between both transitions
remains unclear, a situation that worsens for finite quark chemical
potential, where lattice simulation are on more shaky grounds.

\section{Hadron Resonance Gas Model and Equation of State of QCD}
\label{sec:HRGM}

\begin{figure}[ttt]
\hspace{-0.5cm}
\includegraphics[width=7.3cm,height=5cm]{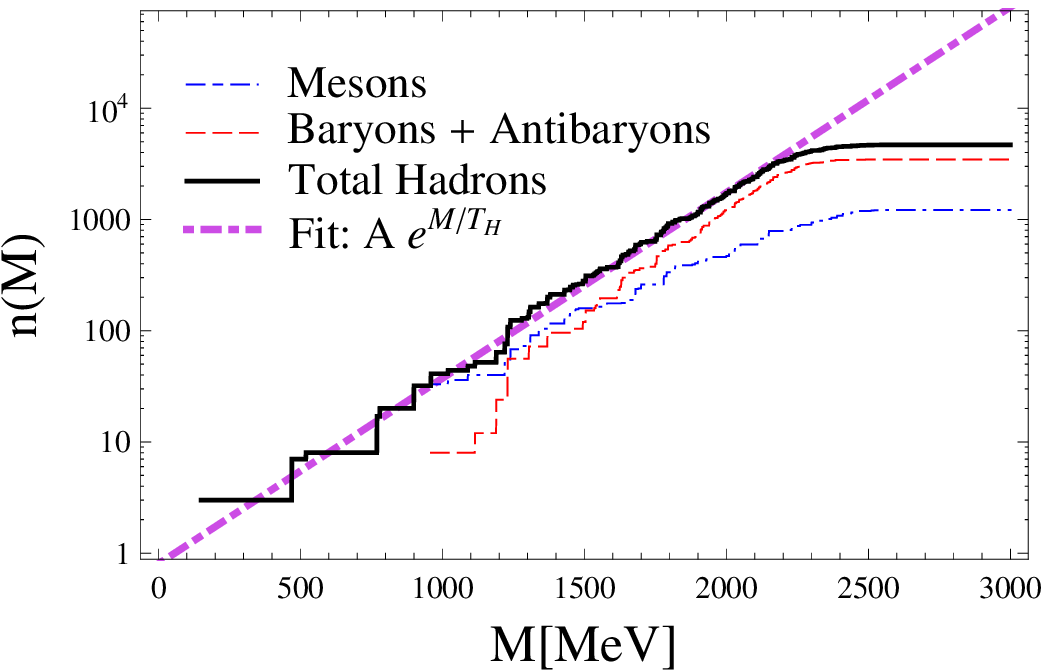} \hfill
\hspace{0.5cm}
\includegraphics[width=7.3cm,height=5cm]{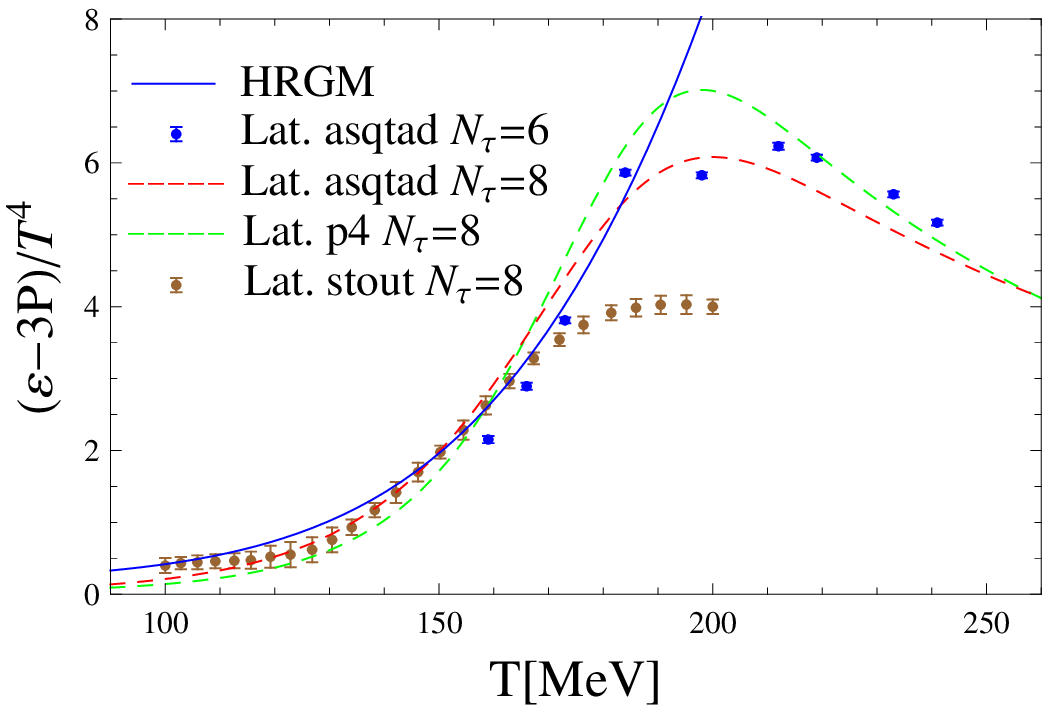}
\caption{Left: Cumulative number $n$ as a function of the hadron mass $M$ (in $\MeV$),
for hadrons with $u$, $d$ and $s$ quarks, computed in the
RQM~\cite{Godfrey:1985xj,Capstick:1986bm}. Right: Trace anomaly $(\epsilon -
3P)/T^4$ as a function of temperature (in MeV). We compare lattice data for
asqtad and p4 actions~\cite{Bazavov:2009zn} (after temperature downshift of
$T_0 = 15\, \MeV$) and stout action~\cite{Borsanyi:2010bp}, with the HRG model
computed with the RQM spectrum shown in the left figure.  }
\label{fig:cumulative_traceanomaly}
\end{figure}

While in the deconfined phase of QCD the quarks and gluons are liberated to
form a plasma, in the confined/chiral symmetry broken phase the relevant
degrees of freedom are hadronic states. The idea of the HRG model is to
describe the equation of state of QCD in terms of a free gas of
hadrons~\cite{Hagedorn:1984hz,Yukalov:1997jk,Cleymans:1999st,Agasian:2001bj,Tawfik:2004sw,
Megias:2009mp,Huovinen:2009yb,Borsanyi:2010cj,Bazavov:2011nk,NoronhaHostler:2012ug},
\begin{equation}
\frac{1}{V}\log Z = -\int \frac{d^3 p}{(2\pi)^3} \sum_\alpha \zeta_\alpha
g_\alpha 
\log \left( 1 - \zeta_\alpha e^{-\sqrt{p^2+M_\alpha^2}/T} \right) 
\,,
\label{eq:hrgm}
\end{equation}
with $g_\alpha$ the degeneracy factor, $\zeta_\alpha=\pm 1$ for bosons and
fermions respectively, and $M_\alpha$ the hadron mass.  The Hagedorn formula
for the trace anomaly follows from Eq.~(\ref{eq:hrgm}) and the standard
thermodynamic relations. It writes
\begin{equation}
\frac{\epsilon - 3P}{T^4} = \sum_{k=1}^\infty \int dM \left( \frac{\partial
n_m(M)}{\partial M} 
+ (-1)^{k+1}\frac{\partial n_b(M)}{\partial M} \right) \frac{1}{2 k \pi^2} 
\left( \frac{M}{T} \right)^3 K_1\left(k \frac{M}{T}\right) 
\,,
\label{eq:traceanomalyHagedorn}
\end{equation} 
where $K_1(z)$ refers to the modified Bessel function.
$n_m$ and $n_b$ are the cumulative numbers of mesons and baryons
(including antibaryons), defined as $n(M) = \sum_\alpha
g_\alpha \Theta( M - M_\alpha )$, where~$\Theta$ is the step
function. $n(M)$ represents the number of hadrons with mass less
than~$M$. Hagedorn proposed that the cumulative number of hadrons in
QCD is approximately given by $n(M) \simeq A \, e^{M/T_H}$, where
$T_H$ is the so called Hagedorn temperature. We show in
Fig.~\ref{fig:cumulative_traceanomaly} (left) the cumulative number of
hadrons with $u$, $d$ and $s$ quarks computed in the Relativized Quark
Model (RQM)~\cite{Godfrey:1985xj,Capstick:1986bm} up to a cutoff
$M\approx 2300\, \MeV$. The total cumulative number can be
approximated to the exponential form with $A = 0.80$, $T_H =
260\, \MeV$ and $\chi^2/\textrm{dof}=0.031$, in the regime $500\,\MeV
< M < 2300\,\MeV$. The spectrum obtained can be used to compute the
trace anomaly within the HRG approach given by
Eq.~(\ref{eq:traceanomalyHagedorn}). The result and its comparison
with lattice data shows that the HRG model gives a good description of
the trace anomaly for $T < 180\,\MeV$, see
Fig.~\ref{fig:cumulative_traceanomaly} (right).

\section{Hadron Resonance Gas Model for the Polyakov loop}
\label{sec:Polyakov_loop}

The expectation value of the Polyakov loop in the irrepresentation $\mu$ of
the color gauge group $\SU(N_c)$ is given by
\begin{equation}
 L_{{\rm QCD},\mu}(T)
:=
\frac{Z_{{\rm QCD},\mu}}{Z_{\rm QCD}}
\,,
\label{eq:L_qcd}
\end{equation}
where $Z_{{\rm QCD}}$ is the physical partition function obtained by
projecting onto states which are singlet at every point, and $Z_{{\rm
QCD},\mu}$ is the partition function with a static color charge in the irrep
$\mu$.  In the confined phase the static source (heavy quark or gluon) is
screened by dynamical quarks and gluons from the medium to form a heavy hadron
(or glueball). Within the present approach, we retain the confining forces
that give rise to the hadron, but neglect the interaction of this hadron to
other dynamical hadrons present in the resonance gas, i.e. non-confining
forces. This assumption is parallel to that of the HRG model for the partition
function.

\subsection{Results in the fundamental representation}
\label{subsec:results_Lfund}

\begin{figure}[ttt]
\hspace{-0.5cm}
\includegraphics[width=7.3cm,height=5cm]{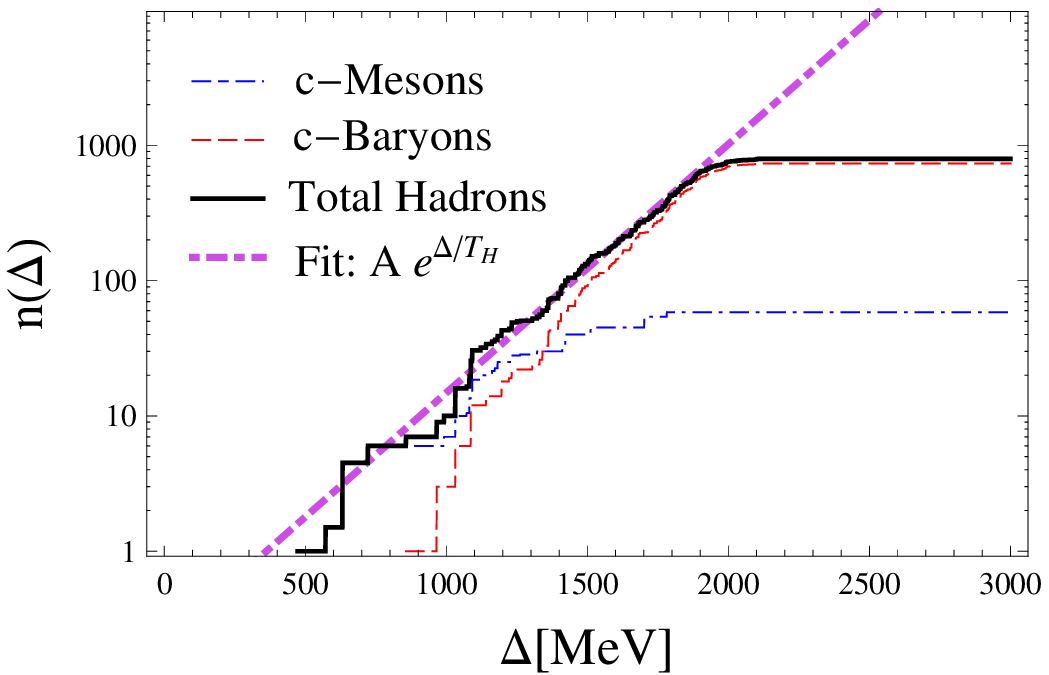} \hfill
\hspace{0.5cm}
\includegraphics[width=7.3cm,height=5cm]{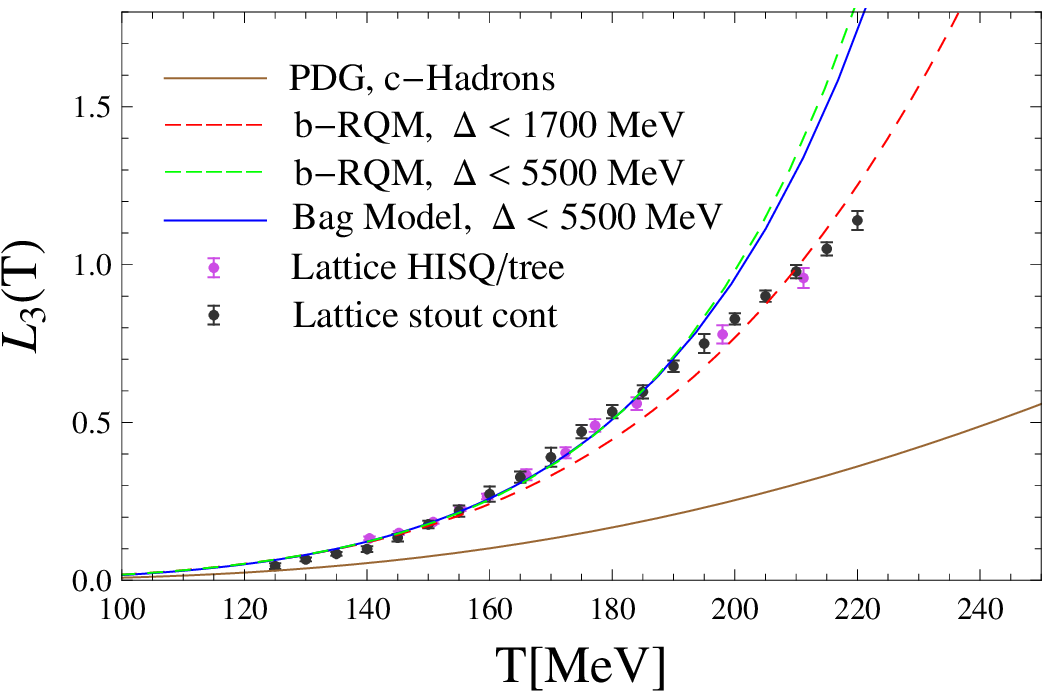}
\caption{
Left: Cumulative number $n$ as a function of the heavy quark subtracted hadron
mass $\Delta = M - m_h$ (in $\MeV$), for the hadron spectrum with a $c$ quark
and light dynamical quarks $u$, $d$ and $s$, computed in the
RQM~\cite{Godfrey:1985xj,Capstick:1986bm}. Right: Polyakov loop in the
fundamental representation as a function of $T$ (in MeV). Lattice data from
the HISQ/tree action~\cite{Bazavov:2011nk} and stout
action~\cite{Borsanyi:2010bp} are displayed. We compare the HRG model result
when including in Eq.~(\ref{eq:hrgm_PL}) the lowest-lying charmed hadrons from
PDG~\cite{Beringer:1900zz}, the RQM spectrum with one $b$ quark and cut-off
$\Delta < 1700 \,\MeV$, $\Delta < 5500 \, \MeV$, and the MIT bag model with
cut-off $\Delta < 5500\, \MeV$~\cite{Megias:2012kb}.  }
\label{fig:rqmbag}
\end{figure}

Based on these considerations, we have shown in~\cite{Megias:2012kb} that a
hadronic representation of the Polyakov loop in the fundamental representation
($\mu = {\bf 3}$) is given by
\begin{equation}
L_{\bf 3}(T) 
\approx \sum_\alpha g_{h\alpha} \, e^{-\Delta_{h\alpha} / T}  \,, 
\qquad 
\Delta_{h\alpha} = M_{h\alpha} - m_h   \,, \label{eq:hrgm_PL}
\end{equation}
where $g_{h\alpha}$ are the degeneracies and $\Delta_{h\alpha}$ are
the masses of hadrons with exactly one heavy quark (the mass of the
heavy quark itself $m_h$ being subtracted).  A natural step is to
check to what extent this hadronic sum rule is fulfilled by
experimental states compiled in the PDG~\cite{Beringer:1900zz}.
Single charmed hadrons are preferable, as there are more available
data than single bottomed hadrons.
Specifically, when considering the lowest-lying single-charmed mesons and
baryons with $u$, $d$, and $s$ as the dynamical flavors, i.e. a total of~12
meson states and 42 baryon states, we find that the result falls short to
saturate the sum rule, cf. Fig.~\ref{fig:rqmbag} (right).  The conclusion is
that one needs many more states than for the trace
anomaly~\cite{Megias:2012kb,Arriola:2013jxa,Bazavov:2013yv}. This suggests using quark model
spectra for $[h\bar q]$ and $[hqq]$ color singlet states with one heavy
$h=c,b$ quark and the remaining light quarks $q=u,d,s$.  This study was
performed in~\cite{Megias:2012kb,Megias:2012hk}. The hadron spectrum obtained
with the RQM is displayed in Fig.~\ref{fig:rqmbag} (left), while we show in
Fig.~\ref{fig:rqmbag} (right) the Polyakov loop computed from
Eq.~(\ref{eq:hrgm_PL}). The result using the MIT bag
model~\cite{Chodos:1974je} is also displayed. Now the sum rule is almost
saturated.

It remains the important question of completeness of hadronic states, and the
possible existence of exotic states as well as their contribution on the HRG
model side of the sum rule. We will give some insights in the next sections.

\subsection{Ambiguities in the hadron resonance gas approach}
  
Within the present approximation, there is some ambiguity as to
exactly which states should be included in the sum rule
Eq.~(\ref{eq:hrgm_PL}). The problem is as follows, let $V$ be the
spatial neighborhood of the static color source with the dynamical
constituents (quarks, antiquarks, gluons). The procedure of just
adding constituents in $V$, to form color singlets with the source,
and computing the resulting spectrum, will certainly produce states
which are spurious. Namely, states composed of a genuine heavy hadron
plus one or more ordinary dynamical hadrons. One can consider two
possible prescriptions to remove the spurious states (see
Ref.~\cite{mrs:prep} for details):
\begin{itemize}
\item Include in the sum rule just configurations of constituents which are
{\em color irreducible}, i.e., without subclusters of constituents forming a color singlet by themselves,
\begin{equation}
L_\mu(T)
:=
\sum_{i, \, {\rm irred}} g_i \, e^{-\beta \Delta_i}
.
\label{eq:L_irred1}
\end{equation}
\item Cancel the reducible configurations by considering the quotient

\begin{equation}
\tilde{L}_\mu(T) := \frac{Z_\mu(T)}{Z_{{\bf 1}}(T)}
,
\qquad \textrm{where} \qquad
Z_\mu(T)
:=
\sum_{i, \,{\rm all}} g_i \, e^{-\beta \Delta_i}
\,.
\label{eq:L_irred2}
\end{equation}
$Z_\mu(T)$ contains all (reducible and irreducible) configurations of type
$\mu$, and $Z_{\bf 1}(T)$ is the physical partition function. 
$Z_\mu(T)$
factorizes in two contributions: the partition function of the neighborhood
$V$ of the static source times a hadron gas with a hole in $V$.
\end{itemize}

Both prescriptions will be studied in the next section within a particular
quark model.  As we will see, the cancellation of reducible configurations
with the denominator in Eq.~(\ref{eq:L_irred2}) only happens for non exotic
states.

\section{Constituent Quark Models and the Hadron Resonance Gas}
\label{sec:Polyakov_cqm}

An effective approach to the physics of QCD at finite temperature is
provided by chiral quark models coupled to gluon fields in the form of
a Polyakov
loop~\cite{Meisinger:1995ih,Fukushima:2003fw,Megias:2004hj,Megias:2005qf,Ratti:2005jh,Megias:2006df,Megias:2006bn,Schaefer:2007pw,Contrera:2007wu,Ghosh:2007wy}. While
most of these works remain within a mean field approximation, we have shown in~\cite{Megias:2004hj} that such approximation erases
information such as the Polyakov loop expectation values in higher
representations. We show in this section that the sum rule for the
fundamental Polyakov loop Eq.~(\ref{eq:hrgm_PL}) is
fulfilled in these models only when one advocates the local and quantum nature of the Polyakov loop. The need of these corrections was stressed in~\cite{Megias:2004hj,Megias:2005qf,Megias:2006df,Megias:2006bn,Megias:2006ke}. Other irreps will be explored~as~well.

\subsection{The Polyakov-Constituent Quark Model}
\label{subsec:Model}

We consider a model that describes QCD using free constituent dynamical quarks and gluons with a Polyakov variable $\Omega(\vx)$ at each point of space. The partition function of the model is given by~\cite{Megias:2004hj,Megias:2006bn,RuizArriola:2012wd,Megias:2013aua}
\begin{equation}
Z_{\rm{PCM}} = \int \prod_{\vx} d\Omega(\vx) \, e^{-S_{\rm{PCM}}(\Omega,T)} \,,
\label{eq:Z_pcm}
\end{equation}
where the matrix $\Omega(\vx)$ is kept as a quantum and local degree of
freedom. $d\Omega(\vx)$ is the invariant $\SU(N_c)$ group integration measure
$(N_c = 3)$ at the point $\vx$. The action splits as a sum of contributions
from each kind of constituent,~and~it~reads
\begin{equation}
S_{\rm{PCM}}(\Omega,T) = \sum_{c = q, \bar{q}, g}  S_c(\Omega,T) \,, \qquad \textrm{where}  \qquad S_c (\Omega,T) = g_c \zeta_c \int \frac{d^3 x d^3 p}{(2\pi)^3} \tr \log \big( 1 - \lambda \zeta_c \Omega_c (\vx) \, e^{-E_c/T}\big)  \,. \label{eq:S_pcm}
\end{equation} 
The action depending on the quarks $S_q + S_{\bar{q}}$ is obtained
from the corresponding fermion determinant, and $S_g$ mimics
gluodynamics \cite{Meisinger:2003id}.  $E_c = \sqrt{\vp^2+M_c^2}$ is
the energy of the constituent $c$, and $M_c$ is the corresponding
constituent mass. The degeneracy factors are $g_{q,\bar{q}}= 2 N_f$
and $g_g = 2$. In the notation of Eq.~(\ref{eq:S_pcm}), $\Omega_c$~is
the Polyakov loop in the fundamental ($c=q$), antifundamental
($c=\bar{q}$) or adjoint ($c=g$) representation, so we can equally use
the standard notation $\Omega_{\bf 3}$, $\Omega_{\bf \bar{3}}$ and
$\Omega_{\bf 8}$ respectively. $\lambda$ is a parameter that counts
the number of constituents. One can always replace $\lambda \to 1$.
After a series expansion in Eq.~(\ref{eq:S_pcm}), the Lagrangian
density for constituent~$c$ reads
\begin{equation}
\cL_c(\vx) = - T g_c \zeta_c \sum_{n=1}^\infty \frac{(\lambda \zeta_c)^n}{n} J_n(M_c,T) \tr  (\Omega_c^n(\vx)) \,, \qquad c = q, \bar{q}, g \,, \label{eq:Lcexpansion}
\end{equation}
(see~\cite{Megias:2013aua,Sasaki:2012bi,mrs:prep} for an alternative
expansion) where we have defined $J_n(M_c,T) := \int \frac{d^3p}{(2\pi)^3} \,
e^{-n E_c/T } 
\sim e^{-n M_c/T}$ for $T \ll M_c$,
displaying the statistical Boltzmann factor characteristic of multi-quark or
multi-gluon states~\cite{Megias:2004hj}. For instance, meson-like
contributions induce corrections of the form $\sim e^{-2M_q/T}$, and
baryon-like contributions behave as $\sim e^{-N_c M_q/T}$. To take into
account quantum corrections in the Polyakov loop, an integration in the color
group must be performed~\cite{Megias:2004hj}.  In addition, there are local
effects associated to the correlation of Polyakov loops.  We assume that the
space is decomposed into domains of size $V_\sigma$, such that two Polyakov
loops are fully correlated if they lie within the same domain and are fully
uncorrelated otherwise. The contribution to the partition function of any such
domain is $\int d\Omega \,e^{-\frac{V_\sigma}{T}\sum_c \cL_c }$. The
volume rule $V_\sigma = 8\pi T^3/\sigma^3$ has been motivated
in~\cite{Megias:2004hj,Megias:2006bn} to describe the crossover, although its
performance at low temperatures remains to be analyzed.

\subsection{Expansion in the number of constituents}
\label{subsec:expansion}

We can perform an expansion of observables in the number of constituents. When
using the estimate given by Eqs.~(\ref{eq:L_irred1}) for the Polyakov loop in
the fundamental representation, one finds within the confined domain
approximation
\begin{eqnarray}
&&\hspace{-1.1cm} L_{{\rm PMC}, \bf 3}  = \lambda \bar{Q}_1 + \lambda^2 \frac{1}{2}  \left( Q_1^2 + Q_2 + 2 G_1 \bar{Q}_1 \right) + \lambda^3 \left( Q_1^2+ \bar{Q}_1 G_1 \right) G_1 \nonumber \\
&&+ \lambda^4 \left[ \frac{1}{4} (5 Q_1^2 G_1^2  - Q_1^2 G_2 - Q_2 G_1^2 + Q_2 G_2 ) + \frac{1}{6} (5 G_1^3 - 3 G_1 G_2 - 2 G_3) \bar{Q}_1  \right]
+ O(\lambda^5) \,, \nonumber \\
&&\hspace{-0.6cm}\simeq  g_q [h\bar{q}] + \frac{g_q}{2}(g_q + 1) [hq^2] + 2g_q [h\bar{q}g] + 2g_q^2[hq^2g] + 4g_q [h\bar{q}g^2] + \frac{g_q}{2}(9 g_q -1)[hq^2g^2] + 4g_q [h\bar{q}g^3]  + {\cal O}(\lambda^5),
\label{eq:L3}
\end{eqnarray}
\begin{figure}[ttt]
\hspace{-0.5cm}
\includegraphics[width=7.3cm,height=5cm]{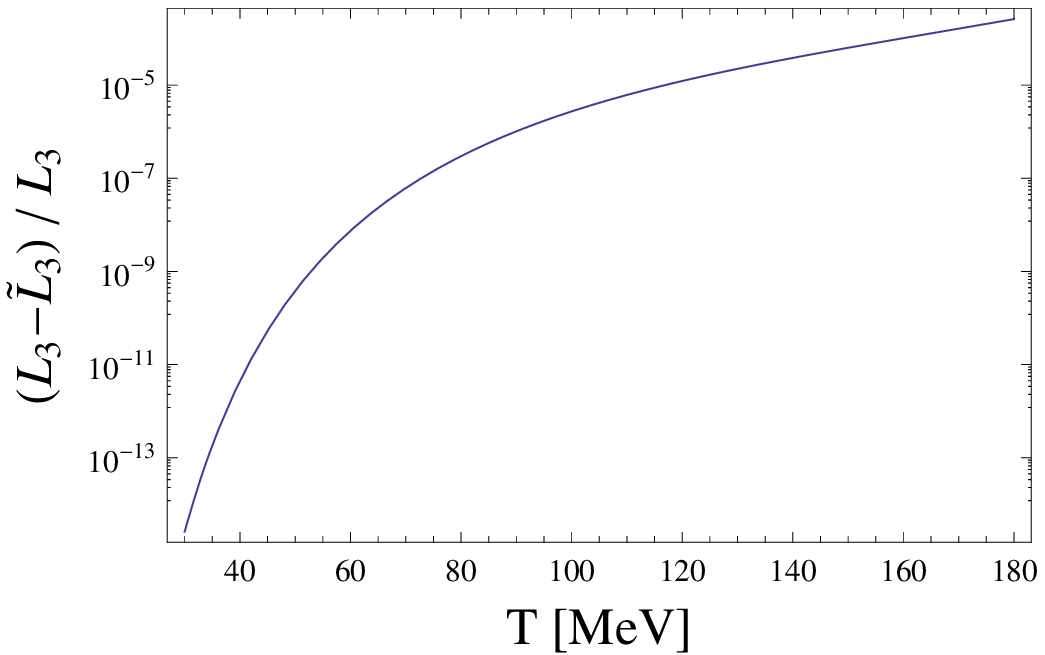} \hfill 
\hspace{0.5cm}
\includegraphics[width=7.3cm,height=5cm]{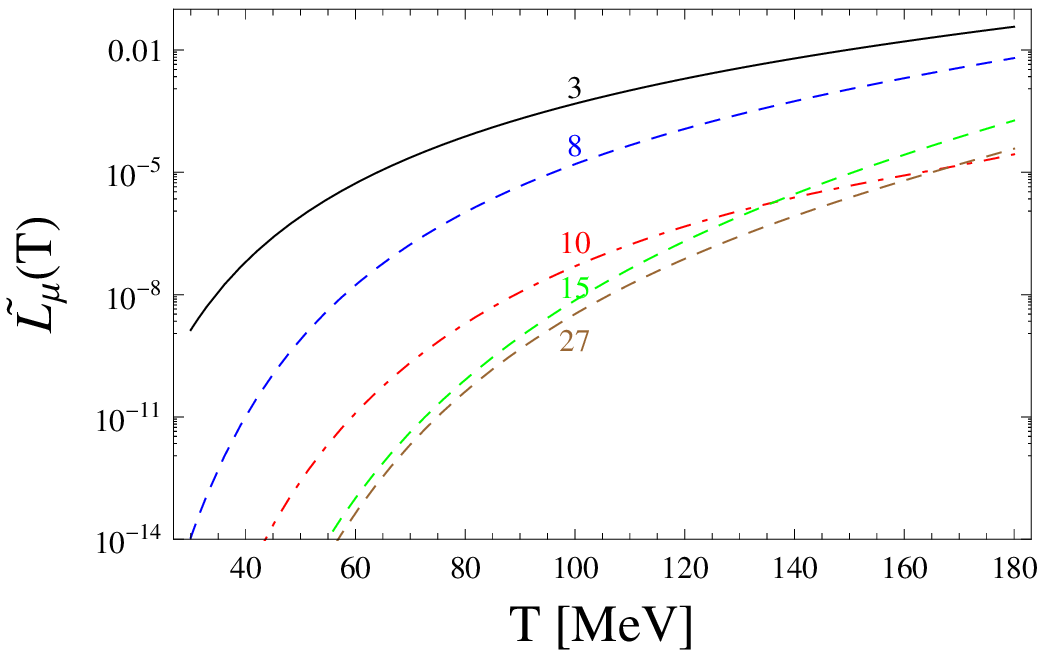} 
\caption{
Left: Difference between $L_{{\rm PCM},{\bf 3}}$ and
  $\tilde{L}_{{\rm PCM},{\bf 3}}$ (normalized to $L_{{\rm PCM},{\bf
  3}}$) as a function of $T$ (in $\MeV$),
  cf.~Eqs.~(\ref{eq:L3})-(\ref{eq:Lt3}). Right: $\tilde{L}_{{\rm
  PCM},\mu}$ as a function of $T$, for several
  irreps. From top to bottom $\mu= {\bf 3}$, ${\bf 8}$, ${\bf 10}$, ${\bf 15}$,
  ${\bf 27}$. In these plots we have included up to four constituents,
  used $N_f=2$, constituent quark and gluon masses $M_q =
  300 \, \MeV$, $M_g = 664 \, \MeV$, and $\sigma = (425\, \MeV)^2$. }
\label{fig:L_pcm}
\end{figure}
where in the second equality we have written the result in a schematic
way. We have defined $Q_n(T) = g_q V_\sigma J_n(M_q,T) \,$ and $G_n(T)
= g_g V_\sigma J_n(M_g,T) \,,$ for quarks and gluons
respectively. $\bar{Q}_n$ is numerically identical to $Q_n$ but it
accounts for $n$ antiquarks. Each factor $Q_n$, $\bar{Q}_n$ or $G_n$
counts as $n$ quarks, antiquarks or gluons, respectively,
corresponding to dynamical constituents from the medium which screen
the Polyakov loop itself, the latter being identified with a heavy
quark source~''$h$''. The factor in front of each term corresponds to
the degeneracy. The first two terms $[h\bar{q}]$ and $[hq^2]$ are
mesons and baryons, respectively, with a heavy quark and one or
several light (anti)quarks. Quantization of the model would produce
the energy levels $\Delta_{h\alpha}$ to be used in
Eq.~(\ref{eq:hrgm_PL}), as explained
in~\cite{RuizArriola:2012wd}. This completes the connection with the
HRG model for the Polyakov loop~\cite{Megias:2012kb}. If the color
cluster decomposition were exact, higher order terms could contain
configurations that can be identified with hybrids, pentaquarks ${\cal
O}(\lambda^5)$, etc.~\footnote{It is noteworthy that tetraquark states
$[h\bar{q}^2q]$ are always reducible. When no confining interactions
are switched off they split into two mesons
$[h\bar{q}][\bar{q}q]$~\cite{mrs:prep}.}  However, the situation is
not so clear as we will see next. When using the second approach,
Eq.~(\ref{eq:L_irred2}), the result one gets is
\begin{equation}
\tilde{L}_{{\rm PMC},\bf 3} \simeq  \cdots - \frac{g_q^2}{6}(g_q+1)(g_q+2)[hq^3\bar{q}] - \frac{g_q}{24}(g_q+1)(g_q+2)(g_q+3)[h\bar{q}^4]  + \frac{g_q}{2}(9 g_q -1)[hq^2g^2] + 4g_q [h\bar{q}g^3] + {\cal O}(\lambda^5) \,,
\label{eq:Lt3}
\end{equation}
where the terms up to ${\cal O}(\lambda^3)$ are identical to those in Eq.~(\ref{eq:L3}). The two approaches differ by terms of ${\cal O}(\lambda^4)$, and some of the configurations in $\tilde{L}_{{\rm PCM},\bf 3}$ at this order appear with negative weights, so the picture is certainly cleaner if just the irreducible configurations are retained. It follows that the ambiguity regarding color singlet clustering affects non-conventional hadrons only. The result for the Polyakov loop in the adjoint representation reads
\begin{eqnarray}
&&\hspace{-2.1cm}\tilde{L}_{{\rm PCM},\bf 8} \hspace{0.1cm} \simeq \hspace{0.1cm} 2[\Omega_{\bf 8} g] + g_q^2 [q\bar{q}]  +  4 [\Omega_{\bf 8}g^2]  + \frac{g_q}{3}(g_q^2-1) \left( [\Omega_{\bf 8}q^3] + [\Omega_{\bf 8}\bar{q}^3] \right) + 4 g_q^2 [\Omega_{\bf 8}q\bar{q}g] + 4 [\Omega_{\bf 8}g^3] +  4 [\Omega_{\bf 8}g^4]  \nonumber \\
&&\hspace{-0.8cm}+ \frac{g_q}{3} (g_q-1)(5 g_q+2) \left( [\Omega_{\bf 8}q^3g] +  [\Omega_{\bf 8}\bar{q}^3g] \right) + 9 g_q^2 [\Omega_{\bf 8}q\bar{q}g^2]   + {\cal O}(\lambda^5) \,,
\end{eqnarray}
 where $\Omega_{\bf 8}$ is an adjoint source at rest, for instance
two heavy quarks coupled adjointwise. In general, up to three constituents the two estimates coincide, 
\begin{equation}
L_{{\rm PCM},\mu} = \tilde{L}_{{\rm PCM},\mu} + O(\lambda^4)\,, 
\qquad
\mu \leq {\bf 27}  \,,
\end{equation}
and they differ by terms of $O(\lambda^5)$ for irreps beyond ${\bf
27}$~\cite{mrs:prep}.  We plot in Fig.~\ref{fig:L_pcm} the Polyakov loop in
several representations, computed with the PCM within the two different
estimates of Eqs.~(\ref{eq:L_irred1}) and (\ref{eq:L_irred2}). We display in
the right figure only those irreps which lead to positive results in the
regime depicted. Negative values are obtained for the representations ${\bf
6}$, ${\bf 15^\prime}$ and ${\bf 24}$ at low temperatures, presumably due to
an unrealistic behavior of $V_\sigma$ in that regime.  A detailed study of
these and other issues will be performed in a forthcoming
paper~\cite{mrs:prep}.

\section{Conclusions}
\label{sec:conclusions}

The thermodynamics of the confined phase of QCD can be described in
terms of a free gas of hadronic states.  A similar approach is
possible for the Polyakov loop in terms of color singlet states with a
static source coupled to dynamical quarks and gluons. We have studied
a particular realization of the HRG description by using a chiral
quark model with Polyakov loop, in which the local and quantum nature
of the Polyakov loop is taken into account. The HRG approach is
however affected by ambiguities when a high enough number of dynamical
constituents are coupled to the static source. A resolution of this
puzzle will shed some light on the existence of exotic states in the
QCD spectrum.


\begin{theacknowledgments}
Supported by Plan Nacional de Altas Energ\'{\i}as (FPA2011-25948), DGI (FIS2011-24149), Junta de Andaluc\'{\i}a grant FQM-225, Spanish Consolider-Ingenio 2010 Programme CPAN (CSD2007-00042), Spanish MINECO's Centro de Excelencia Severo Ochoa Program grant SEV-2012-0234, and the Juan de la Cierva Program.
\end{theacknowledgments}



\bibliographystyle{aipproc}   


\IfFileExists{\jobname.bbl}{}
 {\typeout{}
  \typeout{******************************************}
  \typeout{** Please run "bibtex \jobname" to optain}
  \typeout{** the bibliography and then re-run LaTeX}
  \typeout{** twice to fix the references!}
  \typeout{******************************************}
  \typeout{}
 }



%
%
%
%
%

\end{document}